\documentclass[
reprint,
showpacs,
nofootinbib,
superscriptaddress,
longbibliography,
amsmath,amssymb,prd
]{revtex4-1}

\usepackage{graphicx}  
\usepackage{pstricks}
\usepackage{dcolumn}  
\usepackage{bm}  
\usepackage{esdiff}  
\usepackage{hyperref}  
\usepackage[mathlines]{lineno}  
\usepackage{footnote}

\DeclareMathOperator{\img}{img}

\begin{document}

\title{General Relativistic Geodesy with Clocks on Ground and in Space}
\author{Dennis Philipp}
\author{Jan P. Hackstein}
\author{Eva Hackmann}
\affiliation{Center of Applied Space Technology and Microgravity (ZARM), University of Bremen, 28359 Bremen, Germany}
\author{Claus L\"ammerzahl}
\affiliation{Center of Applied Space Technology and Microgravity (ZARM), University of Bremen, 28359 Bremen, Germany}
\affiliation{Gauss-Olbers-Center, University of Bremen, 28359 Bremen, Germany}
\affiliation{Institute for Theoretical Physics, University of Bremen, 28334 Bremen, Germany}

\begin{abstract}
One of the main tasks of geodesy is to determine the gravity field of the Earth from measurements performed on ground and in space.
High-precision clock comparison has the potential to provide a new observable for the global determination of the Earth's gravitational potential through the gravitational redshift. 
Towards such a global chronometric geodesy, we derive a general relativistic framework beyond the post-Newtonian level for the relativistic redshift and the timing between observers in stationary spacetimes. These observers are equipped with standard clocks and may move along arbitrary worldlines. The redshift is factorized into geometric, transversal-Doppler, and longitudinal-Doppler contributions, providing a transparent separation of gravitational and kinematic effects.
We then show how redshift measurements involving clocks on ground and/or in space can, in principle, be used to determine the mass multipole moments of the underlying spacetime. In the weak field limit, 
our results recover the corresponding Newtonian and post-Newtonian expressions. 
The framework is illustrated for different exact vacuum spacetimes and provides a theoretical basis for using clock comparisons as an additional observable for gravity-field recovery and future relativistic geodesy missions.
\end{abstract}

\pacs{}  
\keywords{}  
\maketitle



\section{Introduction}

One main task of geodesy is to determine the gravity field and orientation of the Earth. 
The conventional way to achieve this is by using gravimeters, gradiometers, and gyroscopes, as well as astronomical observations. 
In Newtonian geodesy, the Earth's geopotential is the fundamental concept and uniquely describes the gravity field.
General Relativity (GR) opens a new perspective on the matter.
There are more gravitational degrees of freedom, and the Newtonian potential is insuffient and needs to be generalized.
However, GR also offers clocks as fundamental measurement devices to probe spacetime geometry, i.e., to determine the gravity field in the context of chronometry.
Significant progress in the development of highly stable and accurate clocks allows us to leverage the general relativistic effect of the gravitational redshift to probe gravitational fields.
This has been proposed for the first time by Bjerhammar \cite{Bjerhammar:1985}, and due to the present stability of atomic clocks of about $10^{-18}$ \cite{Mehlstaubler:2018saw} one can resolve the Newtonian gravitational potential differences at the level of one centimeter. 
Because the comparison of clocks with optical fibers is at least one order of magnitude better than that with free-space links \cite{Droste:2013,Lindvall2025}, one may determine the gravitational potential even on large scales \cite{Lindvall2025,Wu2018, Denker2018,Lisdat2016}.
However, for a global determination, satellites must be used as transmitters or relays, equipped with stable clocks. 
Such clocks onboard satellites can be used to disseminate time and frequency around the world, or they can be used as devices to determine the gravitational potential in space. 
The development of the theoretical foundation of the latter is the purpose of this article, for which the formalism of GR is employed without resorting to post-Newtonian or other approximations. 

In Refs.\ \cite{Philipp:2017, Philipp:2020} a general relativistic formulation, that is, a genuinely chronometric formulation of the gravitational field of the Earth, is given.
It is mainly based on the assumption that the field (in the mean) is stationary and that the laboratory clocks do not move with respect to the rotating Earth\footnote{The worldlines of these clocks, together with the worldlines of all the Earth's constituents, form an isometric (i.e.\ Killing) congruence.}. 
The formalism is used to define a fully general relativistic geoid, beyond post-Newtonian approaches, and it is shown that the equipotential surfaces defined by clock comparisons coincide with those defined by usual gravimeters and gradiometers such as falling corner cubes or atom interferometers.
Weak field results such as those in Ref.\ \cite{Soffel:1987} are recovered in the limit and we also point the reader to Refs.\ \cite{Kopeikin:2015, Kopeikin:2018} for comparison and application.

For chronometry with satellites, we have to extend the formalism to allow for the integration of moving clocks. 
This will be achieved in the present article. 
The comparison of clocks on Earth and satellites via electromagnetic signals propagating through the Earth's atmosphere is not as stable as the comparison done through optical fibers connecting clocks on the ground. 
The stability is generally around $10^{-15}$ and not better than $10^{-16}$. For instance, the currently ongoing ACES mission aims to connect highly precise space clocks with clocks on Earth through an advanced time-transfer system, with objectives including fundamental tests and relativistic geodesy \cite{Cacciapuoti2020fna}. 
However, to carry out chronometry in space, it might be advantageous to pairwise compare clocks in space directly, determine the gravitational potential differences, and deduce global characteristics.

In the first part of the paper, the main theoretical notions are introduced and the problem is defined.
In the second part, general results for space- and ground-based chronometry are derived. 
This includes the discussion of clocks on satellites, moving on geodesics or on arbitrary worldlines, as well as clocks on ground that are located on, e.g., the relativistic geoid. 
It is shown how redshift measurements can be used to determine level surfaces of a relativistic gravity potential \cite{Laemmerzahl:2024,Philipp:2020}, which, in general, depends on the multipole moments of the underlying spacetime. 
The last part contains a brief discussion of the results in special, analytically given spacetimes for illustration.

The assumption of stationarity is made in the following sections. 
For the Earth system, this assumption certainly holds to a high degree because the energy loss associated with gravitational wave emission can be neglected.
All additional dynamics, i.e., time-dependent phenomena can subsequently be included in an adiabatic treatment that assigns a time dependence to parameters in the framework.

\section{Status of clock comparison}

Clock comparison is governed by Special Relativity (SR) and GR. 
SR leads to the Doppler effect (longitudinal and transverse), and GR adds the gravitational redshift and the gravitational time delay. 
The Doppler effect describes the dependence of the difference in measured ticking rates of clocks on relative motion, and the gravitational redshift introduces the dependence on their respective positions in the gravitational field. 
The time delay is an effect due to the propagation of signals in the curved spacetime geometry. 
In general, at least two clocks must be linked to determine their mutual redshift either by optical fiber links or by exchanging electromagnetic signals in free space. 

The first experiment to test the gravitational redshift was done by Pound and Rebka in 1960 \cite{Pound1960}. 
The authors verified the change of a photon's frequency during propagation in the gravitational field of the Earth. 
In 1966, the GEOS-1 satellite was used to observe relativistic Doppler effects \cite{Jenkins1969}. 
In 1976, the famous Gravity Probe A (GPA) experiment was conducted \cite{Vessot1979,Vessot1980,Vessot1989}. 
Two hydrogen masers, one of which was carried to a height of about $10^4\,$km by a Scout D rocket, were compared. 
To quote from the results of this seminal experiment \cite{Vessot1980}: 
\emph{
	``The agreement of the observed relativistic frequency shift with prediction is at the $70 \times 10^{-6}$ level.''
}
The gravitational redshift, however, was confirmed with an accuracy of $1.4 \times 10^{-4}$, see Ref.\ \cite{Vessot1989}.
At present, the best result and highest accuracy of the gravitational redshift is achieved by analyzing clock data from the Galileo satellites 5 and 6, which were launched into elliptic orbits due to a malfunction of the rocket's upper stage. 
This coincidence reduced the usefulness of the satellites for navigation but made them ideal tools for GR tests.
Data analysis and integration over (at least) one year improved the GPA limit to about $4 \times 10^{-5}$ \cite{Hermann:2018,Delva:2018ilu}. 
The Atomic Clock Ensemble in Space (ACES) currently in orbit  proposes to test the gravitational redshift at the $10^{-6}$ level \cite{ACES2011,Cacciapuoti2009,Savalle:2019isy,Cacciapuoti2020fna}, with the aim of another order of magnitude of improved accuracy.

In the past, several satellite-based tests of SR and GR that involve space-based clocks have been proposed, see, e.g., \cite{Laemmerzahl2001, Dittus2007}. 
A recent approach to test the GR prediction of the gravitational redshift uses the RadioAstron satellite resulted in a bound comparable to the GPA experiment \cite{Nunes2023}, and the authors propose future highly elliptic satellite missions that could reach the $10^{-7}$ regime. 
Further prospects for future satellite missions using spacecraft clocks are discussed in, e.g., \cite{Angelli2014} and \cite{Altschul:2014lua}.

For a general overview of ''The Confrontation between General Relativity and Experiment``, we refer the reader to the living review article in Ref.\ \cite{Will:2014}.
Details on relativistic effects in the Global Positioning System (GPS), in particular on satellite timing and Earth-to-satellite redshift, can be found in the review \cite{Ashby2003}, and references therein. 
Moreover, Refs.\ \cite{Kopeikin:2011book} and \cite{Soffel:2019book} are useful reference for applications in various aspects and contain valuable information on, e.g., reference systems and post-Newtonian expansions.
For time transfer in the vicinity of the Earth and general definitions of different time scales, we refer to \cite{Nelson2011}.

Relativistic redshift effects can also serve to test different theories of gravity \cite{Bondarescu2015, Jetzer2017}. 
For example, scalar-tensor theories and their parametrized post-Newtonian form were considered in Ref.\ \cite{Schaerer2014}, in which the authors consider the difference to the GR redshift signals for artificial satellite orbits around the Earth.

All the achievements and considerations mentioned so far (usually) only take into account the first-order post-Newtonian approximation of the relativistic redshift, which contains the Doppler effect and the gravitational redshift to order $\mathcal{O}\big(1/c^2\big)$. 
This may be sufficient to meet the present technological capabilities and to provide the theoretical framework for clocks in space with contemporary accuracy for frequency transfer related to modern IAU and IERS conventions. 
However, to construct a fully relativistic framework for future high-precision experiments in a top-down approach and to obtain a thorough understanding and interpretation of measurement results, we find it necessary to investigate all related notions in GR without relativistic approximations. 
Thus, in this work, we formulate exact general relativistic expressions for chronometry in stationary spacetimes. 
On the basis of the results, weak-field approximations in, e.g., the post-Newtonian framework can be derived.
While stationarity can be regarded as a strong assumption, it is important to note that this assumption holds true in the exterior of the Earth to a high level of accuracy and for a significant duration, spanning typically multiple orbital periods. 
Any dynamics can be subsequently included in the analysis in an adiabatic manner, and time-variable gravity signals can be detected with dedicated (local) ground-based clock networks \cite{Vincent2023} or even low-Earth orbiters \cite{Shabanloui2023}.
Note that chronometry is closely related to the notion of the so-called clock compass \cite{Puetzfeld2018}, which is a theoretical construction for measuring the components of the curvature tensor by clock comparisons in specific constellations.


\section{Setting}
\subsection{Geometry, notation and redshift definition}
We use Einstein's summation convention, Greek indices are spacetime indices and run from 0 to 3, whereas Latin indices are purely spatial indices running from 1 to 3. 
Our metric signature convention is $(-,+,+,+)$. 
If not noted otherwise, we use the framework of Einstein's GR and assume the vacuum field equation to be fulfilled. 

As an illustrative example, the comparison of clocks in the spacetime surrounding our Earth is considered throughout this work. 
However, it is crucial to emphasize that the results of such comparisons are universally applicable to any other stationary situation.

Any time-dependence of the gravitational field can be subsequently included in the analysis in an adiabatic manner by just assuming multipole moments to be time-dependent as usual.

Let ${(\mathcal{M},g)}$ be the relativistic spacetime manifold with metric $g$, for which we assume stationarity, and let $\nabla$ be the Levi-Civita covariant derivative of $g$.

A light signal (null geodesic or optical fiber-guided) is described by a null curve $\lambda$ on $\mathcal{M}$,
\begin{align}
	\lambda: \mathbb{R} &\to \mathcal{M} \notag \\	
	s &\mapsto \lambda(s) \, ,
\end{align}
where $s$ is an affine parameter along the curve. 
The tangent vector to the worldline is denoted by 
\begin{align}
l := \mathrm{d}\lambda / \mathrm{d}s \, \quad \text{with } \quad g(l,l) = 0 \, ,
\end{align}
and it is related to the wave covector $k$ via 
\begin{align}
    l = g^{-1}(k,\cdot) \, ,
\end{align}
where $g^{-1}$ is the inverse metric.

The worldline of an observer is the image of a map $\gamma$,
\begin{align}
	\gamma: \mathbb{R} &\to \mathcal{M} \notag \\	
	\tau &\mapsto \gamma(\tau) \, ,
\end{align}
and we impose the normalization
\begin{align}
	g(u,u) = -c^2 \, , \quad u := \dot{\gamma} := \mathrm{d}\gamma / \mathrm{d}\tau \, , 
\end{align}
that holds along the worldline for the future-pointing tangent $u$.
The curve parameter $\tau$ is then called the proper time.
Clocks that show proper time along their worldlines are called standard clocks. 
They can be operationally characterized as shown in \cite{Perlick:1987}.

In a chart $(\mathcal{U},x)$, with domain $U \subset \mathcal{M}$ and coordinate map $x:\mathcal{M}\to \mathbb{R}^4$, the components are
\begin{subequations}
\begin{align}
	l^\mu &= \dfrac{\mathrm{d}(x\circ \lambda)^\mu}{\mathrm{d}s} \, , \quad k_\mu = g_{\mu\nu} l^\nu \, , \quad u^\mu = \dfrac{\mathrm{d}(x\circ \gamma)^\mu}{\mathrm{d}\tau} \, .
\end{align}
\end{subequations}
and the elapsed proper time along a clock's worldline is\footnote{The four-dimensional worldline length is the elapsed proper time according to the clock postulate.} 
\begin{align}
	\tau = \dfrac{1}{c} \int_{\gamma} \sqrt{g_{\mu\nu}\mathrm{d}x^\mu \mathrm{d}x^\nu} \, .
\end{align}

A given observer associates a momentary frequency\footnote{The frequency is defined up to a multiplicative constant related to the energy at infinity.} $\nu_p$ with a light ray at the event at which their worldlines intersect, ${p \in \mathcal{M} := \img (\gamma) \cap \img (\lambda)}$, where $\img(\cdot)$ denotes a map's image. 
The frequency of a photon at such an event is the map
\begin{align}
    \label{Eq:frequDef}
	\nu_p: T_p\mathcal{M} \supset \mathcal{A}_p  &\to \mathbb{R} \notag \\
	u &\mapsto \nu_p(u) := k(u) \big|_p \, .
\end{align}
Here, $\mathcal{A}_p$ is the set of all timelike, future-pointing, and normalized four-velocities in $T_p\mathcal{M}$. 
Hence, the frequency of the photon is not determined until an observer is chosen.
At $p$, the respective curve parameters are $s_p$ and $\tau_p$, such that
\begin{align}
	\lambda (s_p) := p =: \gamma(\tau_p)\, .
\end{align}
The affine parameter can always be chosen such that ${s_p = 0}$ and we can scale ${\tau_p = 0}$ as well for a single event of interest (offset the clock's dial).
For simplicity, the coordinate time $t_p := t|_p$ can be set to zero. 

\begin{figure}
	\centering
	\includegraphics[width=0.41\textwidth]{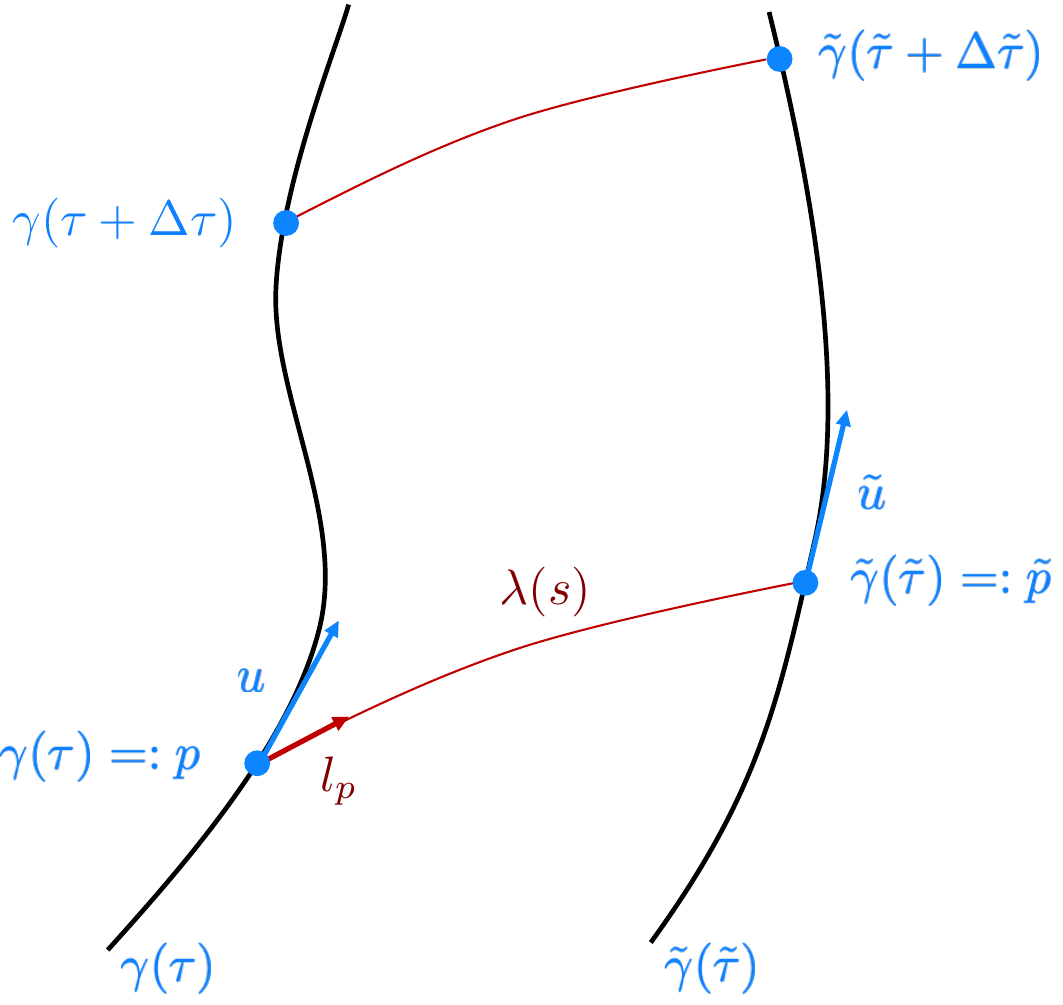}
	\caption{\label{Fig:redshiftSketch} A sketch of the transmission of a signal between the emitter and the receiver for the definition of the redshift. Two worldlines $\gamma$ and $\tilde{\gamma}$ with tangent vectors $u$ and $\tilde{u}$, respectively, are given. The successive exchange of lightlike signals gives rise to the definition of the redshift $z$ according to Eq.\ \eqref{Eq:redshiftDef1}.}
\end{figure}

Now, consider (at least) two distinct observers whose worldlines intersect the same null geodesic.
Let them be described by their respective normalized four-velocities $u$ and $\tilde{u}$, with worldlines (integral curves) $\gamma(\tau)$ and $\tilde{\gamma}(\tilde{\tau})$. 
The redshift $z$ between them is operationally defined by the quotient of emission and reception time increments;
\begin{align}
	\label{Eq:redshiftDef1}
	z+1 := \underset{{\Delta \tau \to 0}}{\mathrm{lim}}  \dfrac{\Delta \tilde{\tau}}{\Delta \tau} = \dfrac{d \tilde{\tau}}{d \tau} \, ,
\end{align}
see the sketch in Fig.\ \ref{Fig:redshiftSketch}.
In GR, there is a universal formula for the redshift in Eq.\ \eqref{Eq:redshiftDef1}, which is \cite{Kermack:1934}
\begin{align}
	\label{Eq:redshiftGR}
	z
	&= \dfrac{\nu_p(u)}{\nu_{\tilde{p}}(\tilde{u})} - 1
	= \dfrac{k(u)\big|_{p}}{k(\tilde{u})\big|_{\tilde{p}}} - 1 \, ,
\end{align}
where 
\begin{enumerate}
	\item $k$ is the wave covector field of the null geodesic $\lambda$ that intersects the observers' worldlines,
	\item $p = \img(\gamma) \cap \img(\lambda)$,
	\item $\tilde{p} = \img(\tilde{\gamma}) \cap \img(\lambda)$.
\end{enumerate}
Note that no metric is needed to define the redshift, and the affine parameter along the light ray can always be scaled such that
\begin{align}
\label{Eq:lightRayStartEnd}
	 \lambda (s_p = 0) = p = \gamma(\tau_p) \, ,
	 \quad \lambda(s_{\tilde{p}} = 1)  = \tilde{p} = \tilde{\gamma}(\tilde{\tau}_{\tilde{p}})\, .
\end{align}
Hence, the affine parameter runs from 0 to 1 between emission and reception.
Note that for geodesic signals, $\tilde{p}$ is on the (future) light cone of $p$ and we restrict all considerations to situations in which there is exactly one null geodesic that connects the two events.
This is certainly true for objects like the Earth and even for more compact Neutron Stars, but not in the immediate vicinity of objects like black holes that have a photon sphere.


\subsection{Emitter--observer maps}
The three quantities $u,\tilde{u},k$ are, in general, not independent. 
For two observers $\gamma, \tilde{\gamma}$, we may choose $p \in \text{img}(\gamma)$ as an emission event. 
Then, this choice (uniquely) determines $k$ and $\tilde{p}$ such that the null geodesic starting at $p$ with initial condition $k|_p$ intersects $\img (\tilde{\gamma})$ at $\tilde{p}$.
Thus, the reception event is a function of the emission event, that is, $\tilde{p} = \tilde{p}(p)$.
The redshift may then be parameterized by either the time of emission or the time of reception, and the emitter-observer problem (EOP) is hidden here implicitly. 
It consists of finding the unique light ray $\lambda$ that satisfies Eq. \eqref{Eq:lightRayStartEnd} for given observer worldlines and a chosen emission event.
No analytic solution to this problem is known (yet), even in simple spacetimes, but there exist analytical approximation schemes \cite{Semerak:2015} and numerical solutions \cite{Hackstein2025}.

\begin{figure}
	\centering
	\includegraphics[width = 0.45\textwidth]{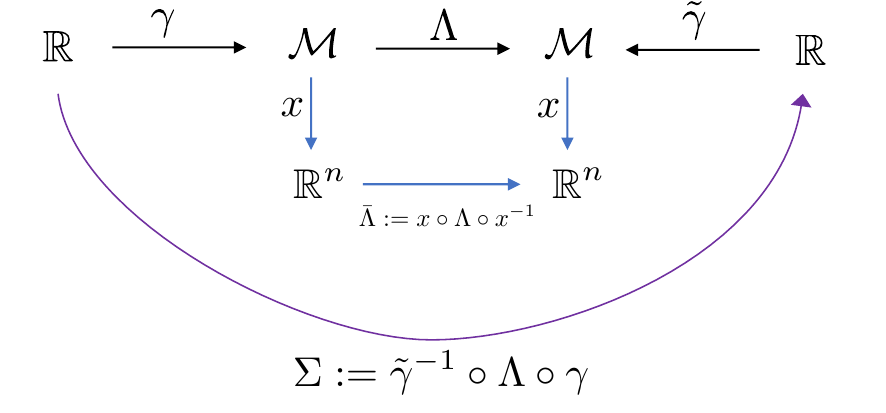}
	\caption{\label{Fig:emitter-observer-map}The emitter-observer maps definitions.}
\end{figure}

However, it is instructive to view the redshift from a slightly different perspective as a map from $\mathbb{R} \to \mathbb{R}$ in the following way.
Let 
\begin{align}
	\Lambda : \text{img} (\gamma) &\to \text{img}(\tilde{\gamma}) \notag \\
	p &\mapsto \Lambda(p) := h^{l_p}_{1}(p) = \tilde{p}
\end{align}
be called the emitter--observer (EO) event map, such that Eq.\ \eqref{Eq:lightRayStartEnd} holds with a connecting light ray $\lambda$, which has the tangent $l_p = g^{-1}(k,\cdot)|_p$ at $p$.
$\Lambda$ maps $p$ to $\tilde{p}$ along the (geodesic) flow of $l_p$ for unit parameter distance\footnote{The unit flow of a vector field $X$ maps a point $p$ to a point $\tilde{p}$ which is on the integral curve of $X|_p$ a unit parameter distance away. Here we use the notation $h^X_d$ to describe the flow in the direction $X$ with parameter distance $d$.}. 
In a chart, of which the domain covers the experiment, $\Lambda$ has a representative 
\begin{align}
    \bar{\Lambda} = x \circ \Lambda \circ x^{-1} : \mathbb{R}^4 \supset x(\img(\gamma))  &\to x(\img(\tilde{\gamma})) \subset \mathbb{R}^4 \notag \\ 
    x(p) &\mapsto x(\tilde{p}) \, .
\end{align}
Because $\gamma$ and $\tilde{\gamma}$ are both curves on ${\mathcal{M}}$, we can also define an EO timing map $\Sigma$,
\begin{align}
	\label{Eq:DefSigma}
	\Sigma := \tilde{\gamma}^{-1} \circ \Lambda \circ \gamma : \mathbb{R} \supset \text{preimg}(\gamma)  &\to \text{preimg}(\tilde{\gamma}) \subset \mathbb{R} \notag \\
	\tau &\mapsto \Sigma(\tau) := \tilde{\tau} \, ,
\end{align}
see Fig. \ref{Fig:emitter-observer-map}.
Here, $\tau$ is the proper time read off the clock on $\gamma$ at the emission of the signal and $\Sigma(\tau) = \tilde{\tau}$ is the proper time read off the clock on $\tilde{\gamma}$ at reception of the signal.
Since for a map from ${\mathbb{R} \to \mathbb{R}}$ we clearly understand the derivative, we can now define the redshift as a function of the emission time by
\begin{align}
	\label{Eq:redshiftDef3}
	z:= \Sigma'-1: \mathbb{R} &\to \mathbb{R} \notag \\
	\tau &\mapsto z(\tau) := \Sigma'(\tau) - 1 \, ,
\end{align}
where the prime denotes the derivative w.r.t. $\tau$.
Therefore, all that is needed to calculate the redshift is, in principle, the map $\Lambda$ and the EOP is encoded in it.
If this map is known (numerically or by an approximate description), the redshift can be easily calculated.

The EO timing map $\Lambda$ is also related to the time transfer function $\mathcal{T}$ in a chart $(\mathcal{U},x)$ with $(x^\mu) = (t,x^i)$ by
\begin{align}
	\label{Eq:TT-function}
	\mathcal{T}(p, \tilde{p}) :&= t(\tilde{p}) - t(p) = t(\Lambda(p)) - t(p) \notag \\
 &= (t \circ \Lambda) (p) - t(p) = \bar{\Lambda}^t (x(p)) - t(p) \, .
\end{align}
Hence, the chart-dependent time transfer function is included in the chart components of $\Lambda$.

However, the total redshift will in general contain contributions from the geometry of the spacetime and from the Doppler effect known from SR due to relative motion.
The redshift equation\eqref{Eq:redshiftGR} can now be rewritten as
\begin{align}
	\label{Eq:redshiftResult}
	z(\tau) = \dfrac{k(u) \big|_p}{k(\Lambda^*(u)) \big|_{\Lambda(p)}} -1 \, ,
\end{align}
where $u = u(\tau)$ and $\Lambda^*$ is the push-forward defined w.r.t.\ the map $\Lambda$ \footnote{The push forward map $\Lambda^*$ is defined such that ${\Lambda^*(u) (f) = u (f \circ \Lambda) \, , \, \forall f \in C^\infty(\mathcal{M,\mathbb{R}}), \, u \in \mathcal{TM}}$, i.e., for all smooth and real-valued functions.}.
This notation clearly demonstrates that the reception event is a function of the emission event, effectively encapsulating the essence of the EOP.
Note that tangent vectors of curves are pushed forward to tangent vectors of the image curves, which are constructed by the underlying flow, that is, $\tilde{u} = \Lambda^*(u)$. 

In the chart, the redshift can then be decomposed into
\begin{align}
	\label{Eq:redshiftChart}
	z(\tau)+1 &= \dfrac{k_t(0) u^t + k_i u^i \big|_{p}}{k_t(1) \tilde{u}^t + k_i \tilde{u}^i \big|_{\Lambda(p)}} \notag \\
	&= \dfrac{u^t(\tau) + \dfrac{k_i(0)}{k_t} u^i(\tau)}{\tilde{u}^t \big( \Sigma(\tau) \big) + \dfrac{k_i(1)}{k_t} \tilde{u}^i \big( \Sigma(\tau) \big)} 
\end{align}
because in a stationary spacetime and in adapted coordinates $k_t$ is a constant of motion along the null geodesic. 
The wave covector $k$ at emission and reception is given by the respective derivative of the time transfer function,
\begin{align}
	\dfrac{k_i(0)}{k_t} = c \dfrac{\partial \mathcal{T}(p,\tilde{p})}{\partial x^i} \, , \quad \dfrac{k_i(1)}{k_t} = -c \dfrac{\partial \mathcal{T}(p,\tilde{p})}{\partial \tilde{x}^i} \, ,
\end{align}
Thus, the redshift can also be written as 
\begin{align}
	\label{Eq:redshift-TTF}
	z(\tau)+1 = \dfrac{u^t(\tau) + c \dfrac{\partial \mathcal{T}(p,\tilde{p})}{\partial x^i} u^i(\tau)}{\tilde{u}^t \big( \Sigma(\tau) \big) -c \dfrac{\partial \mathcal{T}(p,\tilde{p})}{\partial \tilde{x}^i} \tilde{u}^i \big( \Sigma(\tau) \big)} \, ,
\end{align}
which demonstrates, again, in combination with Eqs.\ \eqref{Eq:DefSigma} and \eqref{Eq:TT-function} that all the information is, in principle, contained in the EO map $\Lambda$. 
Hence, to model the redshift between given observer worldlines, several equivalent expressions are available as outlined above.

In studies which analyze the redshift of satellites with respect to Earth-bound clocks  \cite{Hermann:2018,Delva:2018ilu}, usually clock residuals instead of redshifts are investigated. 
This can be done by relating the proper times $\tau$ and $\tilde{\tau}$ using Eq.\ \eqref{Eq:redshiftDef1}.
Thus,
\begin{align}
	\mathrm{d} \tilde{\tau} = \mathrm{d}\tau \, (z+1) \, ,
\end{align} 
and, therefore,
\begin{align}
    \tilde{\tau} = \Sigma(\tau) = \tau + \int_{\gamma} z(\tau) \, \mathrm{d}\tau + \mathcal{C} \, ,
\end{align}
with some constant offset $\mathcal{C}$.
Measured clock residuals along a satellite's orbit can then be used to calculate the redshift, which is compared to models for tests of GR and other theories of gravity. 

In the Newtonian limit, we have $\Sigma(\tau) = \tau + \mathcal{C}$ and, thus, $z = \Sigma' = 0$ as time is absolute and there is no redshift.

For two stationary observers in a stationary spacetime we have\footnote{Actually the result holds for any two observers who are members of a Killing congruence.}
\begin{align}
	\Sigma(\tau) = \mathcal{C}_1(p,\tilde{p}) \, \tau + \mathcal{C}_0 \, .
\end{align}
Thus, the redshift is constant but position dependent and given by
\begin{align}
	\label{Eq:redshiftConstant}
	z = \mathcal{C}_1(p,\tilde{p}) \, ,
\end{align}
which relates to the existence of a time-independent redshift potential, see eq.~\eqref{Eq:redshiftStationaryLimit} below.

A precise description of the involved EOP can now be formulated as follows.
Given observers $\gamma$ and $\tilde{\gamma}$, the EOP consist in finding for every event $p$ on the emitter's worldline a null vector $l_p$ such that
\begin{align}
    \img(\tilde{\gamma}) \equiv \left\lbrace \tilde{p} = h^{l_p}_{1} (p) \, \, \forall p \in \text{img}(\gamma) \right\rbrace \, .
\end{align}

The most difficult step in the theoretical redshift modeling is to solve the EOP.
To this end, we can consider some kind of approximate or intermediate modeling. A numerical solution was implemented in the software package GREOPy \cite{Hackstein2025}. On the post-Newtonian level, an approximate analytical approach was developed in \cite{Schlake:2025btk}. For very strong gravitational fields the approximations reviewed in \cite{Semerak:2015} might be useful. Then, once the EOP is solved, the redshift can be calculated accordingly.


\subsection{Metric, observers, and velocities}
Consider two observers as introduced above and let them exchange light signals. 
So far, neither of them needs to be geodesic or distinguished in any way.

A general stationary spacetime can be described by the metric
\begin{align}
	\label{Eq:stationaryMetric}
	g = g_{tt} \mathrm{d}t^2 + 2 g_{ti} \mathrm{d}t \mathrm{d}x^i + g_{ij} \mathrm{d} x^i \mathrm{d} x^j,
\end{align}
where we use coordinates $(t,x^i)$ and assume that the time-coordinate $t$ is adapted to the symmetry such that $\partial_t$ is the timelike Killing vector field.
Then, we can write 
\begin{align}
g_{tt} = -c^2 e^{2\phi}
\end{align}
with the gravitational redshift potential $\phi$\footnote{This potential is also called the gravitoelectric potential.}, which is constant along the trajectory of stationary observers \cite{Ehlers:1993, Hasse:1988} and which is the basis for so-called gravitoelectric effects and derived notions, e.g., the definition of the relativistic geoid and heights \cite{Philipp:2017, Philipp:2020}. 
Far away from the gravitating source, we have $g_{tt} \to -c^2$. 
The level surfaces of $\phi$ are called isochronometric, i.e., two stationary clocks (clocks on $t-$lines) on the same level surface have a vanishing redshift \cite{Philipp:2017, Philipp:2020}. 
Alternatively, the metric can also be written as
\begin{align}
	g = -e^{2\phi} \left(c\mathrm{d}t + \sigma_i \mathrm{d}x^i \right)^2 + e^{-2\phi} \gamma_{ij}\, \mathrm{d}x^i \mathrm{d}x^j \, ,
    \label{Eq:stat_metric}
\end{align}
and $\gamma_{ij}$ is the spatial metric seen by the Killing observers.
All gravitational degrees of freedom are encoded in $\phi$, $\sigma_i$, and $\gamma_{ij}$.
The gravitomagnetic degrees of freedom, $\sigma_i$, can further be related to the twist potential $\omega$,
\begin{align}
   \partial_i \omega = -e^{2\phi} \epsilon_{ijk} D^j \sigma^k \, ,
\end{align}
where $D$ is the covariant derivative w.r.t.\ $\gamma_{ij}$.
The twist potential conceptualizes a gravity degree of freedom that is not present in Newtonian geodesy.
Its level surfaces can be studied via, e.g., Sagnac interferometers and allow to obtain information about rotational contributions to the gravity field \cite{Laemmerzahl:2024}.
In total, there are ten degrees of freedom, but note also that four can be eliminated by coordinate transformations.

For Killing observers at a coordinate position ${x := (x^i)}$, the relation between proper time and coordinate time is
\begin{align}
	\mathrm{d}t = e^{-\phi(x)} \mathrm{d}\tau \quad  \Leftrightarrow \quad u^t_\text{Killing} = e^{-\phi(x)} \, .
\end{align}
Thus, these stationary observers are redshifted by $\exp(\phi(x))$ w.r.t\ observers at infinity.

From the normalization of an arbitrary observer's four-velocity, we have
\begin{align}
	\label{Eq:normalization1}
	-c^2 = g_{tt} \big( u^t \big)^2 + 2 g_{ti} u^t u^i + g_{ij} u^i u^j \, ,
\end{align}
and only the spatial velocity components are, in principle, subject to measurement. 
We can express $u^t$ as a function of $u^i$,\footnote{We choose the positive solution since $t$ and $\tau$ shall both increase.}
\begin{align}
	u^t = -\dfrac{\sigma_i u^i}{c} + e^{-\phi} \sqrt{e^{2\phi} \left(\dfrac{\sigma_i u^i}{c} \right)^2 + 1 + \dfrac{u^2}{c^2}} \, ,
\end{align}
where $u^2 := g_{ij} u^i u^j$.
In the static limit ($g_{ti} \to 0$), this reduces to
\begin{align}
	u^t = e^{-\phi} \sqrt{1 + u^2 /c^2 } \, .
\end{align}
Let the coordinate velocity components $v^i\in\mathbb{R}$ be defined by
\begin{align}
	v^i := \dfrac{\mathrm{d}x^i}{\mathrm{d}t} \quad \Rightarrow \quad u^i = v^i\, u^t \, ,
\end{align}
and, thus, the normalization leads to
\begin{align}
	\label{Eq:u^t_of_v^i}
	u^t = \dfrac{ e^{-\phi} }{\sqrt{1 + \dfrac{2\sigma_i v^i}{c} - e^{-2\phi} \dfrac{v^2}{c^2}}} \, ,
\end{align}
where $v^2 := g_{ij} v^i v^j$.
In the static limit, $u^t$ is given by
\begin{align}
	u^t = \dfrac{ e^{-\phi} }{\sqrt{1 - e^{-2\phi} \dfrac{v^2}{c^2} }} \, .
\end{align}
Note that the above equation resembles a modified Lorentz factor and in the special relativistic limit $\exp(-\phi) \to 1$.

We can introduce yet another velocity $w$ with spatial components $w^i$, which is defined by a stationary standard clock at an observer's momentary position $x$
\begin{align}
	w^i := \dfrac{\mathrm{d}x^i}{\mathrm{d}\tau_s} = \dfrac{\mathrm{d}x^i}{\mathrm{d}t} \dfrac{\mathrm{d}t}{\mathrm{d}\tau_s} = v^i e^{-\phi(x)}\, .
\end{align}
Here $\tau_s$ is the proper time on the locally stationary standard clock at $x$
and we observe that
\begin{align}
	\phi(x) = -\log \left( \dfrac{w^i}{v^i} \right) \, .
\end{align}
Thus, observing the velocity of a satellite on the two time scales $t$ and $\tau_s$ allows, in principle, to determine the metric function $\phi(x)$ locally.
Note that the equation involves the fraction of vector components labeled by $i$. It holds $\forall i = 1,2,3$  and yields a scalar quantity indeed.
However, the coordinate time scale $t$ is unfortunately not accessible directly in experiments.

In total, we have three different velocities with components $u^i, v^i, w^i$, related to three different time scales. They serve to simplify equations, take another view on the problem, and are beneficial for specific applications, respectively.

\section{Gravity level surfaces: chronometry}
One application of redshift measurements is to determine level sets (i.e., equipontential surfaces) of a relativistic gravity potential $U^*$,
\begin{align}
    U^* := c^2 \left( e^\phi -1 \right) \, ,
\end{align}
for stationary spacetimes.
It is derived from the redshift potential $\phi$ and also realizes the relativistic geoid in terms of a particular level surface \cite{Philipp:2020}.
Note that, depending on the choice of (co-rotating) coordinates, $U^*$ can include centrifugal effects.
This potential is of great importance for chronometry and relativistic geodesy in general.
The level surfaces of $\phi$ are also the level surfaces of $U^*$ and in the weak field limit $U^* \to W$, where $W$ is the Newtonian gravity potential (geopotential) used in conventional geodesy.

The determination of level surfaces of $U^*$ allows us to infer the spacetime's multipole moment structure and, thus, the underlying geometry.
The mass moments are important for, e.g., height references, global unified height systems, as well as navigation and positioning. 
This adds yet another -- chronometric -- data channel to conventional gravity field recovery and experiments such as GOCE and GRACE, which determine the Earth's (Newtonian) gravity field, but the approach here is entirely formulated within GR.

In the special case that one of the two observers, say $\tilde{\gamma}$, is stationary, i.e., at rest on a $t$-line, we have 
\begin{align} 
	1 + z(\tau) 
	= e^{\phi_0} \left(u^t(\tau) + \dfrac{k_i(0)}{k_t} u^i(\tau) \right) \, ,
\end{align} 
where $\phi_0 = \phi|_{\tilde{p}}$ is the value of the redshift potential at the stationary observer's position. 
We may think of an Earth-bound clock positioned on the geoid such that the adapted coordinate system is corotating with the Earth.
The equation above can be used to theoretically model the redshift between an Earth-bound clock and a clock onboard a moving satellite.

The well-known limit for two stationary observers, e.g., two clocks on the Earth's surface, follows immediately,
\begin{align}
	\label{Eq:redshiftStationaryLimit}
1 + z = \dfrac{u^t}{\tilde{u}^t} = e^{\phi|_{\tilde{p}} - \phi|_p} =: e^{\Delta \phi} \, ,
\end{align}
which is the constant $\mathcal{C}_1$ in \eqref{Eq:redshiftConstant}. 
This equation is the basis for modeling the comparison of any two stationary Earth-bound clocks linked by, e.g., optical fiber networks \cite{Mehlstaubler:2018saw,Philipp:2017}.
It establishes chronometry on Earth because it allows us to infer level surfaces of $\phi$ from the comparison of two clocks in a (global) network.
If all these clocks were adjusted to produce a vanishing pairwise redshift, they would be isochronometric, that is, on the same level surfaces of the relativistic gravity potential $U^*$.

If the velocities $u, \tilde{u}$ are known for moving clocks, analytical models of the redshift can be constructed, given the solution to the EOP. 
Using the normalization \eqref{Eq:normalization1} to eliminate $u^t$ and $\tilde{u}^t$ in the redshift equation \eqref{Eq:redshiftChart} allows to write the solution as a function of the spatial velocities.
In the static limit the redshift can be given in a compact form,
\begin{align}
	1 + z 
	= \dfrac{ e^{-\phi(x)} \sqrt{1 + u(\tau)^2/c^2} + \dfrac{k_i(0) u^i(\tau)}{k_t} }{e^{-\phi(\tilde{x})} \sqrt{1 + \tilde{u}(\Sigma(\tau))^2/c^2} + \dfrac{k_i(1) \tilde{u}^i(\Sigma(\tau))}{k_t} } \, .
\end{align}
The stationary version looks more complicated but follows straightforwardly from \eqref{Eq:normalization1} and \eqref{Eq:redshiftChart}.

However, the four-velocities $u, \tilde{u}$ are not measured in real observations anyway.
Rather, coordinate velocities $v^i$ may be used to proceed.
Then, the coordinate time $t$ parameterizes the redshift and the formula neatly factorizes into
\begin{align}
	\label{Eq:redshiftGeneral1}
	1 + z(t) 
	= \dfrac{u^t(t)}{\tilde{u}^t(\tilde{t})} \dfrac{\left( 1 + \dfrac{k_i(0) v^i(t)}{k_t} \right)}{\left( 1 + \dfrac{k_i(1) \tilde{v}^i(\tilde{t})}{k_t} \right)} \, ,
\end{align}
where the signal reception time is ${\tilde{t} = \bar{\Lambda}^t(x(t))}$ and ${x(t) := x \circ \gamma(t)}$ is the emission event in the chart with emission time $t$.
The second factor in \eqref{Eq:redshiftGeneral1} describes the longitudinal (linear) part of the Doppler effect. 
It is proportional to the line-of-sight projection of the observers' coordinate velocities and will be abbreviated as follows.
\begin{align}
	\mathcal{D}(t,\tilde{t}) := \dfrac{ 1 + \dfrac{k_i(0)}{k_t} v^i(t)  }{ 1  + \dfrac{k_j(1)}{k_t} \tilde{v}^j(\tilde{t}) } \, .
\end{align}
The first factor in \eqref{Eq:redshiftGeneral1} contains the gravitational (geometric) contribution and the transversal part of the Doppler effect.
Inserting $u^t$ and $\tilde{u}^t$ from Eq.\ \eqref{Eq:u^t_of_v^i} leads to 
\begin{multline}
	\label{Eq:redshiftGeneral2}
	1 + z(t) \\
	= e^{\Delta \phi} \sqrt{\dfrac{2\sigma_i(\tilde{x}) \tilde{v}^i(\tilde{t}) /c + 1 - e^{-2\phi(\tilde{x})} \tilde{v}^2(\tilde{t})/c^2}{2\sigma_j(x) v^j(t) /c + 1 - e^{-2\phi(x)} v^2(t)/c^2}} \, \mathcal{D}(t,\tilde{t}) \, ,
\end{multline}
from which the static limit follows,
\begin{align}
	1 + z 
	= e^{\Delta \phi} \dfrac{\sqrt{1 - e^{-2\phi(\tilde{x})} \dfrac{\tilde{v}(\tilde{t})^2}{c^2} }}{ \sqrt{1 - e^{-2\phi(x)} \dfrac{v(t)^2}{c^2} } } \, \mathcal{D}(t,\tilde{t}) \, .
\end{align}
Here, $\Delta \phi = \phi(\tilde{x})-\phi(x)$.
It is a product of three factors with clear interpretation each (geometry, transversal Doppler, longitudinal Doppler).
Note that in the stationary case the transversal Doppler contribution is mainly modified by gravito-magnetic contributions coming from $\sigma_i$.

Expression \eqref{Eq:redshiftGeneral2} is invariant under the transformation $\phi \to \phi - C$, with some constant $C$.
This means that only potential differences contribute to the gravitational redshift.
One way to see this is to apply the transformation at the level of the metric and re-define coordinate time to absorb the effect, i.e., 
\begin{align}
	\phi \to \bar{\phi} := \phi - C \quad \Rightarrow \quad \mathrm{d}T := e^{C} \mathrm{d}t \, .
\end{align}
Thereupon, we have 
\begin{align}
	v^i = \dfrac{\mathrm{d}x^i}{\mathrm{d}t} \to V^i = \dfrac{\mathrm{d}x^i}{\mathrm{d}T} = v^i e^{-C} \, , \quad g_{ti} \to g_{Ti} = g_{ti} e^{C} \, .
\end{align}

Now we rewrite the redshift \eqref{Eq:redshiftGeneral2} in the coordinates $(T,x^i)$ and find the same result as in the coordinates $(t,x^i)$. 
Hence, there is a gauge freedom in the potential $\phi$, which is related to time translations.
The weak-field limit fixes the gauge as done for the potential in Newtonian gravity.

\subsection{Limits}
In the post-Newtonian (pN) limit, the redshift \eqref{Eq:redshiftGeneral2} reduces to
\begin{align}
	1 + z =\left( 1 + \dfrac{\Delta U}{c^2} + \dfrac{v^2-\tilde{v}^2}{2c^2} \right) \dfrac{1+\vec{n}(x)\cdot \vec{v}/c}{1+ \vec{n}(\tilde{x}) \cdot \vec{\tilde{v}}/c } \, ,
\end{align}
where $U$ is the Newtonian gravitational potential and $\vec{n}$ is the local line-of-sight vector (the direction of photon reception $\vec{k}=-|\vec{k}|\vec{n}$), see \cite{Will:2014}. 
Hence, in the pN approximation at $\mathcal{O}(c^{-2})$ there are contributions due to the
\begin{itemize}
	\item positions in the gravitational field: $\dfrac{\Delta U}{c^2}$
	\item state of motion (transversal Doppler): $\dfrac{v^2-\tilde{v}^2}{2c^2}$
	\item signal exchange (longitudinal Doppler): $\dfrac{1+\vec{n}\cdot \vec{v}/c}{1+ \vec{n} \cdot \vec{\tilde{v}}/c }$
\end{itemize}
In the theory of GR, however, the (different) contributions are not simply separable but can be factorized to some extent only as shown above in eq.~\eqref{Eq:redshiftGeneral2}.

A local tetrad $\{e_{\hat{\mu}}\}$ can be introduced at the respective observer's position,
\begin{align}
	e_{\hat{t}} = u \, , \quad k_{\hat{\mu}} = e_{\hat{\mu}}{}^\nu k_\nu
\end{align}
All indices and symbols indicated with a hat are tetrad-related.
In the tetrad, the redshift is
\begin{align}
	1 + z = \dfrac{ \hat{k}(x) + \dfrac{k_{\hat{n}}(x) v^{\hat{n}}(x)}{c} }{ \hat{k}(\tilde{x}) \,\sqrt{ 1-\hat{v}(x)^2/c^2}} \, ,
\end{align}
where 
\begin{align}
	\hat{k} = \sqrt{ \delta^{\hat{m}\hat{n}} k_{\hat{m}} k_{\hat{n}} } \, , \quad \hat{v} = \sqrt{ \delta_{\hat{m}\hat{n}} v^{\hat{m}} v^{\hat{n}} } \, , \quad v^{\hat{m}} = u^{\hat{m}} u^{\hat{t}} \, .
\end{align}
From this expression, the special relativistic limit follows easily. Note that the gravitational contribution to the redshift is included in the spatial length of the wave covector, as can be seen in the limit of no relative motion.

In the special relativistic limit, we obtain the usual relation
\begin{align}
	1 + z = \dfrac{1+\vec{n}(x)\cdot \vec{v}/c}{\sqrt{1-\dfrac{v^2}{c^2}}} \, ,
\end{align}
in the rest system of one observer, i.e., $v$ is the relative velocity.
In the pure Newtonian limit $(c \to \infty)$, it follows that $z \to 0$ since Newtonian time is absolute. 

In the following we apply the general results to special constellations, in which at least one of the two observers is stationary or geodesic. 
The main goal is to determine the relativistic gravity potential $U^*$ (or, equivalently, the redshift potential $\phi$) in terms of its level surfaces.

\subsection{Two geodesic observers}
\label{sec:ACMC}
Consider two satellites moving along geodesics, both of which are equipped with standard clocks. 
For each geodesic observer, we have a conserved specific energy, 
\begin{align}
	\label{Eq:geodeticEnergy1}
	-E := u_t = g_{tt} u^t + g_{ti} u^i = u^t \big( g_{tt} + g_{ti} v^i \big) \, .
\end{align}
Thus, 
\begin{align}
	\label{Eq:geodeticEnergy2}
	u^t = \dfrac{-E}{g_{tt} + g_{ti} v^i} = \dfrac{e^{-2\phi} E/c^2}{1+ \sigma_i v^i / c} \, ,
\end{align}
and the redshift factorizes into
\begin{align}
	\label{Eq:redshiftDoubleGeodesic}
	1 + z(t) &= e^{2 \Delta \phi} \dfrac{E}{\tilde{E}} 
	\left( \dfrac{1+\sigma_i(\tilde{x}) \tilde{v}^i(\tilde{t})/c}{1+\sigma_i(x) v^i(t)/c} \right) \mathcal{D} \, .
\end{align}
It depends on the level surface differences $\Delta \phi$, on the energy ratio $E / \tilde{E}$, on some Coriolis-like terms induced by the frame dragging, and on the longitudinal Doppler contributions $\mathcal{D}$. 
The static limit is given by
\begin{align}
	1 + z(t) &= e^{2 \Delta \phi} \dfrac{E}{\tilde{E}} \mathcal{D} \, .
\end{align}
From \eqref{Eq:redshiftDoubleGeodesic} we find that the potential differences $\Delta \phi$ can be determined by
\begin{align}
	\label{Eq:redshift2geodesics}
	\Delta \phi = \dfrac{1}{2} \log \left( \dfrac{\tilde{E}}{E} \dfrac{1+\sigma_i(x) v^i(t)/c}{1+\sigma_i(\tilde{x}) \tilde{v}^i(\tilde{t})/c} \dfrac{(z(t)+1)}{\mathcal{D}(t,\tilde{t})} \right) \, .
\end{align}
The energy ratio is constant and the redshift as well as positions, velocities, emission (reception) times, and angles are observables. 
The Doppler contributions have to be handled with care because they are typically orders of magnitude larger than the gravitational redshift. 
Then, $\Delta \phi$ can be deduced for given $\sigma_i$. 
This scheme allows us to calculate how the geodesics intersect the level surfaces of $\phi$. 

We may use a multipolar model in terms of mass and spin moments for $\phi$ and $\sigma_i$, which are then determined as the best fit to the position, velocity and redshift data.

Following Thorne \cite{Thorne1980}, we assume that an ACMC (asymptotically Cartesian and mass centered) coordinate system $(t,x^i)$ is used, and introduce the corresponding mass and spin multipoles. For notational ease, we quote them here exactly as in Thorne's original paper \cite{Thorne1980} only for the $g_{00}$ component, together with an explicit expression up to order $1/r^3$ for both $g_{00}$ and $g_{0j}$ \cite{Guersel1983}; general expressions and an algorithm to compute terms to arbitrary order can be found in Refs.~\cite{Thorne1980,Guersel1983}. The metric functions are then
\begin{subequations}
\label{Eq:multipoleExpansions}
\begin{align}
e^{2\phi} & = 1 - \frac{2m}{r} - \sum_{l=1}^\infty \frac{1}{r^{l+1}} \Big[ \frac{2(2l-1)!!}{l!} m_{A_l} n^{A_l} + \sum_{n=1}^l F_{n-1} \Big] \nonumber\\ 
& = 1 - \frac{2m}{r} + \frac{2m^2}{r^2} - \frac{2m^3}{r^3} - \frac{3 m_{kl} n^k n^l}{r^3} + \mathcal{O}(r^{-4})
\end{align}
and
\begin{multline}
e^{2\phi} \sigma_j = \frac{2 \epsilon_{jkl} S^k n^l}{r^2} \left( 1- \frac{m}{r} \right) + \frac{4 \epsilon_{jkl} S^k_m n^l n^m}{r^3} + \mathcal{O}(r^{-4})\,.
\end{multline}
\end{subequations}
Here $A_l = i_1 \ldots i_l$ is a multi-index, $n^{A_l} = n^{i_1} \ldots n^{i_l}$, $n^a = x^a/r$, $\epsilon_{jkl}$ is the flat Levi-Civita symbol, $m = GM/c^2$ is the mass monopole, $m_{A_l}$ are the mass moments (in units of $(\text{length})^{l+1}$; specifically, $m_{kl}$ is the mass quadrupole), and $S^k$ and $S^k_m$ are the spin moments (in units of $(\text{length})^2$ and $(\text{length})^3$, respectively). The symbol $F_n$ indicates contributions from the n-th pole, that depend on spherical harmonics of order 0 to n but are independent of the radius \cite{Thorne1980,Guersel1983}.
Using expansions like the one above, redshift measurements between the satellites and ground stations allow to determine all the moments included in $\Delta \phi$.
Using \eqref{Eq:redshift2geodesics}, we can sort the right-hand side, which is the measurement data, into an observation vector $\vec{b}$ of which each entry corresponds to one measurement. For the $\sigma_i$ terms, the spin-dipole approximation is used.
For the left-hand side of \eqref{Eq:redshift2geodesics} we construct a model based on the multipole decomposition of $\exp(\Delta \phi)$, which is a function of the moments and positions.
For each measurement, a model vector $\vec{a}(m_{A_l})$ is constructed, where $\{m_{A_l} \}$ is the set of multipole moments.
These are determined by a best fit, e.g., a least square adjustment to 
\begin{align}
	\vec{a}(m_{A_l}) - \vec{b} = 0 \, .
\end{align}
Note that the model constructed here is the basis for chronometry in space with two geodesic clocks, e.g., two drag-free satellites.

The particular cases in which only one observer is stationary are treated in the next section for the determination of the level surfaces of $\phi$ with an Earth-bound stationary clock, but the modeling of mass moments works in a similar way.

The relativistic multipole moments used in this paper are those defined by Thorne. 
It is shown in the available literature that there exists a one-to-one correspondence to the moments defined by i) Beig and Simon and ii) Geroch and Hansen, see Ref.\ \cite{Guersel1983,Quevedo:1990}.
All these definitions are invariant and formulated at the level of General Relativity, i.e., without the need of a post-Newtonian framework.
However, all those moments are defined at infinity. For applications in geodesy, moments related to source integrals are preferred, which is only possible in approximate frameworks.
In this case, the Blanchet-Damour (BD) moments should be used \cite{Blanchet:1998}. See also the fundamental work in the so-called DSX papers \cite{DSXI,DSXII,DSXIII,DSXIV}.
The philosophy of the present paper is to investigate chronometry in GR. 
Thus, for this purpose we consider the multipole expansions above from which also the Newtonian ones are recovered in the limit.

\subsection{Two stationary observers}

Clocks at rest on the Earth can be modeled by stationary observers.
In this case, both are Killing observers and part of the same isometric congruence. 
The redshift between them is particularly simple, see Eq.\ \eqref{Eq:redshiftStationaryLimit}; all Doppler terms vanish and we are left with
\begin{align}
	\Delta\phi = \ln \left(1 + z \right) \, .
\end{align}
Any two stationary observers on the same level surface of $\phi$ measure a vanishing mutual redshift, and thus the EOP is implicitly solved.
In terms of multipole moments in an ACMC coordinate system, the level surface differences can be calculated explicitly from \eqref{Eq:multipoleExpansions}
\begin{widetext}
\begin{align}
	\Delta\phi = \frac{1}{2} \ln \left( 
	\frac{1 - \frac{2m}{\tilde{r}} 
	- \sum_{l=2}^\infty \frac{1}{\tilde{r}^{l+1}} \Big[\frac{2(2l-1)!!}{l!} m_{A_l} n^{A_l} + \sum_{n=1}^l F_{n-1}\Big]}{1 - \frac{2m}{r} 
	- \sum_{l=2}^\infty \frac{1}{ r^{l+1}} \Big[\frac{2(2l-1)!!}{l!} m_{A_l} n^{A_l} + \sum_{n=1}^l F_{n-1}\Big]} \right) \, .
\end{align}
\end{widetext}
For measured observer positions in a pairwise redshift measurement, e.g., in clock networks, the relativistic mass moments $m_{A_l}$ can be determined by a model fit. 
Note that this result is exact and generalizes the commonly shown expression,
\begin{align}
    z = \dfrac{\nu - \tilde{\nu}}{ \tilde{\nu}} = \dfrac{\Delta U}{c^2} \, ,
\end{align}
which is valid to first pN order only.
The equation above can be used for chronometry in Earth-bound clock networks and allows us to determine the relativistic mass moments.

\subsection{Earth to satellite}
Let one of the two observers, say $\tilde{\gamma}$, be a stationary observer. 
We might think again of a clock corotating with the Earth on the surface. 
The coordinates are constructed such that this observer moves on a $t-$line, that is,
\begin{align}
	\big( \tilde{u}^\mu \big) = \big( \tilde{u}^t, 0 \big) \, , \quad 
	\tilde{u}^t = e^{-\phi(\tilde{x})} := e^{-\phi_0} = \text{const.} 
\end{align}
Let $\gamma$ be an arbitrarily moving observer, e.g., a clock on a satellite. 
Their redshift as a function of emission time $t$ is
\begin{multline}
	1 + z(t) 
	= \dfrac{e^{\Delta \phi}\left( 1 + \dfrac{k_i(0)}{k_t} v^i(t) \right)}{\sqrt{1 + \sigma_i(x) v^i(t)/c - e^{-2\phi(x)} v^2(t)/c^2}} \, .
\end{multline} 
We should express all measured observables, such as velocities, and metric functions in the frame associated with the stationary observer on the Earth, who observes  proper time $T$. If the clock is located on the relativistic geoid, 
$T$ can be thought of as International Atomic Time (TAI).
Clocks on the Earth's geoid show no mutual redshift (``they tick at the same rate'') but a clock network cannot consistently be mutually Einstein synchronized because of the nonvanishing twist vector on the rotating Earth. The velocity components $V^i$ of the satellite on the timescale of the Earth-bound observer are 
\begin{align}
	V^i := \dfrac{\mathrm{d}x^i}{\mathrm{d}T} = v^i e^{-\phi_0} \, .
\end{align}
We can use $T$ as a new coordinate, which leads to 
\begin{multline}
	\label{Eq:redshiftStationaryMoving}
	1 + z
	= \dfrac{e^{\Delta \phi}\left( 1 + \dfrac{k_i(0)}{k_T} V^i(T) \right)}{\sqrt{1 + e^{\phi_0}\sigma_i(x) V^i(T) / c - e^{-2\Delta \phi} V^2(T)/c^2}} \, .
\end{multline}
with 
\begin{align}
	k_T := k_t e^{-\phi_0} 
\end{align}
being the energy constant of motion of the light ray w.r.t.\ $T$. 
The redshift is a function of position and velocity, that is, the state vector of the satellite, since it contains observables $(x,V)$ in addition to the direction of the emitted light ray. 
Obviously, the expression is invariant under the scaling $\phi \to \phi + C$ as the dependence on the redshift potential is only via $\Delta \phi$.

If the redshift between the Earth-bound clock and the satellite as well as the satellite's momentary state vector are continuously monitored, level surfaces of the redshift potential can be determined. 
These are the sets $\{ (\tilde{x}^i) \}$ for which \eqref{Eq:redshiftStationaryMoving} holds with the same $\Delta \phi$, i.e., a level surface of $\phi$ is the set $\{ (x^i) \}$ for which 
\begin{equation}	
	\label{Eq:redshiftEarth}
	\dfrac{ (1 + z) \sqrt{1 + e^{\phi_0}\sigma_i(x) \frac{V^i(T)}{c}c - e^{-2\Delta \phi(x)} \frac{V^2(T)}{c^2}} }{ 1 + \dfrac{k_i(0)}{k_T} V^i(T)} = \text{const.}
\end{equation}
If the satellite covers a large region of the Earth's exterior with successive orbits, or if multiple satellites are monitored, then after some time discretized level surfaces in terms of point clouds can be determined.
Again, a model can be set up in terms of mass and spin moments for Eq. \eqref{Eq:redshiftStationaryMoving} to determine the multipoles from the measurement data as described above.
In addition, multiple clocks can be used on the geoid to increase data coverage.
Operationally, each satellite may emit a spherical wave at a given frequency, of which one particular direction reaches the Earth-bound clock. 
It measures the received frequency and computes the redshift. 
Moreover, the satellite sends its momentary position and velocity at signal emission in, e.g., the GNSS framework. 
As GNSS uses a timescale related to TAI, the velocity corresponds to the velocity $(V^i)$ in the equation above and all data to deduce the level surfaces are present in principle.

If we assume a geodesic motion of the satellite, the redshift simplifies to
\begin{align}
	\label{Eq:redshiftEarthGeodesic}
	z + 1 = \dfrac{e^{2\Delta \phi}\dfrac{E_T}{c^2} \left( 1+\dfrac{k_i(0)}{k_T}V^i(T) \right)}{1+ e^{\phi_0}\sigma_i \frac{V^i(T)}{c}} \, .
\end{align}
Here $E_T$ is the energy constant of motion of the satellite on the $T$-timescale.
The velocity components $V^i$ follow from the geodesic equation, and the $k_i(0)$ are related to the direction of the emitted light ray, which must be determined to connect the emitter and receiver; see the EOP above.
The $k_i$ can also be related to emission angles of the light signals on the observer's celestial sphere.
However, as far as redshift models are concerned, an analytical solution for $k_i$ will generally not be possible.

Using the relation \eqref{Eq:multipoleExpansions} for $\sigma_j$ up to $1/r^2$, the redshift potential can now be determined from its level sets. 
For a particular level surface $\{(x^i)\}$ it holds that
\begin{align}
	\dfrac{(z + 1) \left(1 + e^{\phi_0}\sigma_i \frac{V^i(T)}{c}\right)}{\dfrac{E_T}{c^2}\left(1+\dfrac{k_i(0)}{k_T}V^i(T) \right)} = \text{const.}
\end{align}
In combination with the expansions \eqref{Eq:multipoleExpansions} all mass and spin moments can be determined from measurement data.
Thus, via the level surfaces of the potential, of which the geoid is a distinguished one, the potential itself and, thereupon, the multipole moment structure of the spacetime can be determined.

Up until now, we did not discuss the spatial coordinates to be used. 
The only assumption so far is that they are adapted to the Earth-bound observer.
Among the suitable coordinates to be used are, e.g., radar coordinates.
To this end, we consider that on the worldline of the observer $\gamma$, a mirror reflects the light signal.
The reflection is received, and we denote the time of signal emission by $T_1$ and the time of reception of the reflected signal by $T_2$.
Then, we associate a radar time $\Theta$ and a radar distance $R$ with the reflection event by
\begin{align}
	\Theta = \dfrac{1}{2} \left( T_2 + T_1 \right) \, , \quad R = \dfrac{1}{2} \left( T_2 - T_1 \right) \, .
\end{align}
Now, we can also introduce two angular coordinates on the celestial sphere of the observer in the following way.
A local tetrad $\{ e_{(i)} \}$ is attached to the observer $\tilde{\gamma}$ with $e_{(0)} = \tilde{u}$.
Now, the tangent to the reflected light ray is considered in the tetrad system.
This tangent is proportional to
\begin{multline}
	- e_{(0)}(T_2) + \cos \varphi \sin \vartheta \, e_{(1)}(T_2) + \sin \varphi \sin \vartheta \, e_{(2)}(T_2) \\ 
	+ \cos \vartheta \, e_{(3)}(T_2) \, .
\end{multline}
The coordinates of the satellite are $(\Theta,R,\varphi,\vartheta)$ and we can use the spatial coordinates $(R,\varphi,\vartheta)$ to determine the level surfaces of the redshift potential $\phi$. To cover the global level surfaces in the vicinity of the Earth, multiple ground stations must be considered.

\subsection{The energy approach}

Let the observer $\tilde{\gamma}$ be stationary on the Earth's surface, and let the observer $\gamma$ be a geodesic satellite. 
Using the normalization of the satellite's four-velocity as well as the conserved energy \eqref{Eq:geodeticEnergy2}, we get
\begin{align}
	e^{-\Delta \phi} = \dfrac{E_T}{c^2} \dfrac{\sqrt{1+2e^{\phi_0}\sigma_i \frac{V^i}{c}-\frac{V^2}{c^2}}}{1+e^{\phi_0}\sigma_i \frac{V^i}{c}} \, . 
\end{align}
Hence, measurements of the satellite's conserved energy and the velocity in the frame of the observer on Earth also give access to the potential differences $\Delta \phi$. 
The conserved energy can be related to initial conditions, i.e., momentary Keplerian orbital elements. 
The velocity has to be considered as a function of the momentary position so that the level surfaces of $\phi$ can be inferred from \eqref{Eq:geodeticEnergy2}.

The main difference between both approaches to determine the level surfaces of $\phi$ is that the redshift approach is always applicable while the energy approach is only valid for geodesic motion, unless a relativistic model of all perturbations is available. 
However, the energy approach is a viable method in Newtonian mechanics as well \cite{Jekeli:1999}, which is recovered in the weak field limit:
In a static situation, we have
\begin{align}
	\Delta \phi = -\log \left( \dfrac{ E_T }{c^2} \sqrt{1 - \dfrac{V^2}{c^2}} \right) \, ,
\end{align}
so that for level surfaces of $\phi$ we have $\{ (\tilde{x}^i) \}$ for which
\begin{align}
	V^2 = g_{ij}V^i V^j = \text{const.} \, ,
\end{align}
i.e., the squared spatial velocity (speed) must be constant, which is formally equivalent to the Newtonian result.
Note that in a stationary spacetime this relation is modified to
\begin{align}
	\dfrac{\sqrt{1+2e^{\phi_0}\sigma_i \frac{V^i}{c} - \frac{V^2}{c^2}}}{1 + e^{\phi_0}\sigma_i \frac{V^i}{c}} = \text{const.} \, ,
\end{align}
of which the leading-order terms are
\begin{equation}
	1 - \frac{1}{2c^2} \left( V^2 + \left( \sigma_i V^i \right)^2 \right) = \text{const.} \, ,  
\end{equation}
that is, 
\begin{equation}
V^2 + \left( \sigma_i V^i \right)^2 = \text{const.}
\end{equation}
Hence, frame-dragging effects modify the relation by an additional contribution.

Note that the results yield an interesting new interpretation for the level surfaces of $\phi$.
On a level surface it is true that
\begin{itemize}
	\item Two clocks at rest show no mutual redshift.
	\item The acceleration of freely falling bodies is orthogonal to the level surface.
	\item In the Earth-to-satellite clock comparison, the quantity
 \begin{equation}
 E_T \, \dfrac{\sqrt{1+2\sigma_i \frac{V^i}{c} - \frac{V^2}{c^2}}}{1+\sigma_i \frac{V^i}{c}}
\end{equation}
is constant for geodesic orbits with energy $E_T$ and velocity $(V^i)$ crossing the level surface.
\end{itemize}
The equivalence of the first two has been shown in Ref.\ \cite{Philipp:2017}.

\subsection{Mirror tracking}
Let one observer be stationary and corotating with the Earth. 
The second observer is moving and carries a mirror instead of a clock, which is assumed to reflect incoming signals in a perfect way.
Using a two-way Doppler scheme, we can determine the frequency quotient of emitted and received signals (after reflection) as follows.

\begin{figure}
	\includegraphics[width=0.49\textwidth]{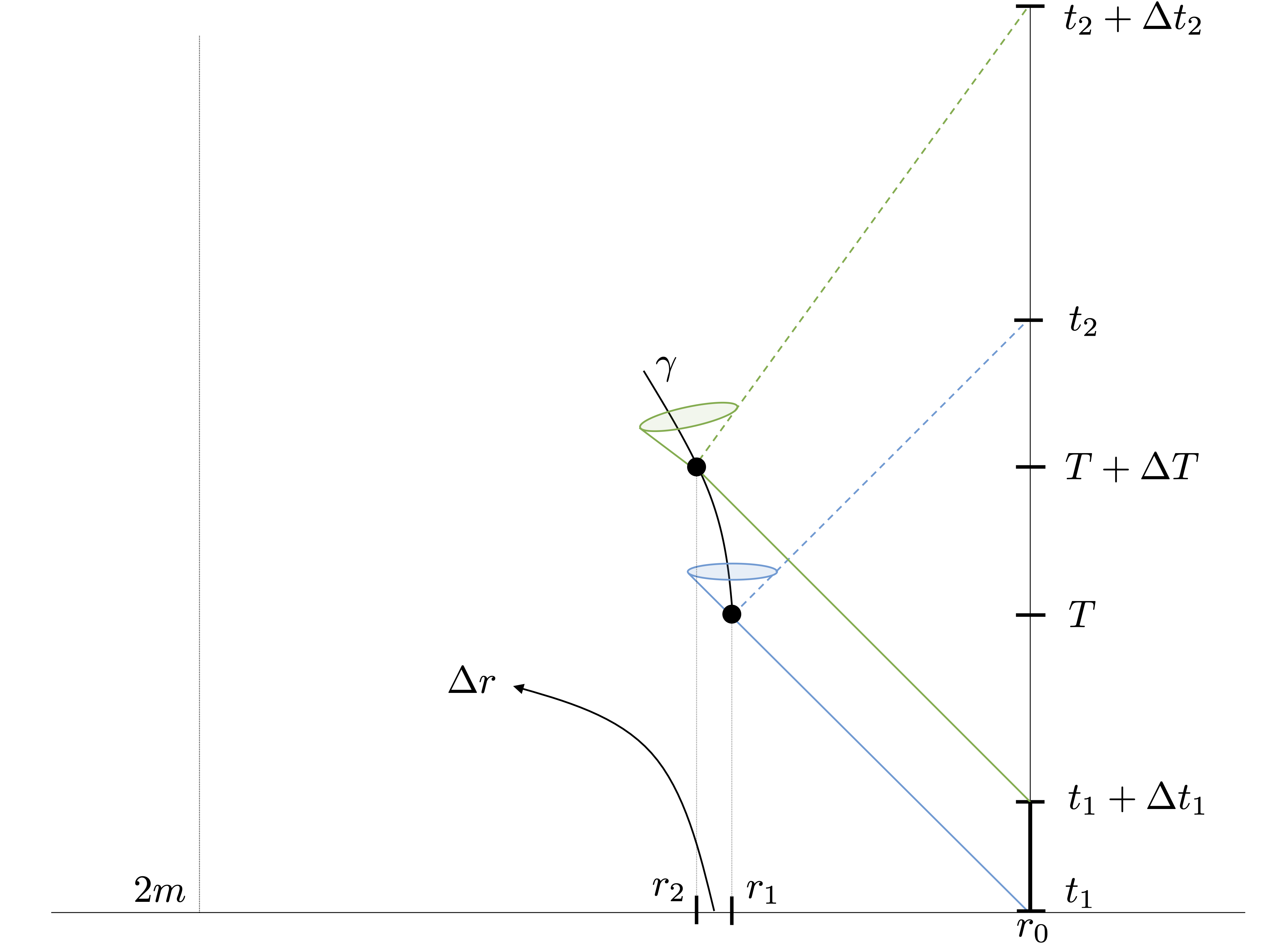}
	\caption{\label{Fig:2wayDoppler} Sketch of two-way Doppler mirror tracking in a Schwarzschild spacetime for radial mirror motion. At smaller radii, light cones tilt more (in these coordinates) causing a contribution to radar time differences.}
\end{figure}

We assume that a signal is emitted at $t_1$ by the first observer and reflected at the mirror such that we assign a radar time $T = (t_2 + t_1)/2$ to the reflection event, where $t_2$ is the time of arrival of the reflected signal. 
A second signal is emitted at $t_1 + \Delta t_1$ and the reflection is received at $t_2 + \Delta t_2$ so that we assign a radar time $T + \Delta T$ to the second reflection event; see Fig.\ \ref{Fig:2wayDoppler}.
For simplicity, we take $t_1=0$. 
Then
\begin{align}
	\Delta T := \dfrac{1}{2} (\Delta t_2 + \Delta t_1) \, .
\end{align}
The observable quotient $\Delta T / \Delta t_1$ leads to quotient of emitted and received frequency,
\begin{align}
	1 + \dfrac{\nu_\text{em}}{\nu_\text{rec}} = 2 \lim_{\Delta t_1 \to 0} \dfrac{\Delta T}{\Delta t_1} \, .
\end{align}
Note that $\Delta t_1$ is the chosen time increment of two successive signals, which is an inverse frequency. 
$\Delta T$ is the observable radar time difference of two successively received reflections. 
It will contain contributions from geometry, intuitively because of tilting light cones, as well as contributions from the velocity of the mirror.
For a light signal, we have that
\begin{multline}
	\dfrac{dt}{ds} = \dfrac{g_{0i}}{2c^2} e^{-2\phi} \dfrac{dx^i}{ds} + \\\sqrt{\left( \dfrac{g_{0i}}{2c^2} e^{-2\phi} \dfrac{dx^i}{ds} \right)^2 + \dfrac{g_{ij}}{c^2} e^{-2\phi} \dfrac{dx^i}{ds} \dfrac{dx^j}{ds}} \, .
\end{multline}
For a static spacetime, we may explicitly rewrite the result as
\begin{align}
	\dfrac{dt}{ds} = \dfrac{e^{-\phi}}{c} |\vec{l}| \, ,
\end{align}
with the tangent to the light ray $\vec{l} = (dx^i/ds)$. 
The travel time of the first light ray is then given by
\begin{align}
	t_2 = 2 \int_1 \dfrac{e^{-\phi(s)}}{c} |\vec{l_1}(s)| ds \, ,
\end{align}
where $\int_1$ denotes that the integral is to be performed along the first light ray's worldline. 
Instead of the coordinate time, we continue to use the proper time of the stationary observer, $\tau = \exp(\phi_0) t$, in which $\phi_0$ is the potential at the position of the observer.
It could be a standard clock on the geoid such that $\tau$ is TAI.
For the increase in radar time, it follows that
\begin{multline}
	\Delta T = \int_2 \dfrac{e^{\Delta \phi}}{c} |\vec{l_2}(s)| ds - \int_1 \dfrac{e^{\Delta \phi}}{c} |\vec{l_1}(s)| ds + \Delta t_1 \, .
\end{multline}
Hence, the measurement of $\Delta T$ is sensitive to $\phi$ in an integral way and an application is shown in the last section.
Note that in general the two integrals will be different if the mirror moves and that a laboratory experiment including freely falling mirrors is covered in principle.
For the frequency quotient, we have
\begin{multline}
    1 + \dfrac{\nu_\text{em}}{\nu_\text{rec}} = 2 \lim_{\Delta t_1 \to 0} \left[ \dfrac{1}{c\Delta t_1} \left( \int_2 e^{\Delta \phi} |\vec{l_2}(s)| ds - \right. \right. \\
   \left. \left. \int_1 e^{\Delta \phi} |\vec{l_1}(s)| ds + \Delta t_1 \right) \right] \, .
\end{multline}
Thus, the result depends on potential differences, i.e., on gravity gradients.
In a stationary spacetime, the integrals are more involved, but the philosophy of the measurement does not change.
To lowest order, we find the known result from Doppler tracking, 
\begin{align}
	1+z = \dfrac{v}{c} \, ,
\end{align}
which will at higher orders be supplemented by corrections from the curvature of spacetime.
The leading-order correction comes from the mass monopole; see the example in the next section.

\section{Spacetime Application}
\label{sec:applications}
For the following applications, we focus on Earth-to-satellite redshift measurement schemes to briefly illustrate the above results.
In general, assuming redshift measurements to be short relative to the temporal variation of the Earth's gravity field, a stationary spacetime metric of the form~\eqref{Eq:stat_metric} may be considered.
Then as described in the previous sections, one can construct the spacetime metric expansion defined by Thorne \cite{Thorne1980} in an ACMC\footnote{De-Donder (harmonic) coordinates are an example of an ACMC coordinate system.} coordinate system $(t, x^i)$ and the spin-dipole approximation, where the gravitational redshift potential $\phi$ and the gravitomagnetic degrees of freedom $\sigma^i$ are given by equations~\eqref{Eq:multipoleExpansions}.
From this, equation~\eqref{Eq:redshiftGeneral2} can be used to model the Earth-satellite redshift, where the mass multipoles can be determined from fitting measurement data as outlined above.
Here, to keep the following discussion of the above mentioned results brief, assume in addition axial symmetry of the underlying spacetime.

In axisymmetric and stationary spacetimes, we may choose coordinates $t, \varphi$ related to the symmetry such that $\partial_\varphi$ is the spacelike Killing vector field of which the orbits are closed circles and $\partial_t$ is the Killing vector field related to time translations. 
The remaining two coordinates are labeled $(x^1, x^2)$. 
We may write the metric as
\begin{align}
	g = -c^2 e^{2\phi} \mathrm{d}t^2 + 2 g_{t\varphi} \mathrm{d}t \mathrm{d}\varphi + g_{AB} \mathrm{d} x^A \mathrm{d} x^B + g_{\varphi\varphi} \mathrm{d}\varphi^2 \, ,
\end{align}
where $A,B = 1,2$. 
There are constants of motion along a light ray and along the worldline of a geodesic observer, respectively,
\begin{align}
	&(k_t, k_\varphi) \, , \quad  (E := -u_t, L := u_\varphi) \, .
\end{align}
The remaining free functions are $u^1,u^2$ and $k_1,k_2$, which need to be determined from the geodesic equations for light and massive particles, respectively. 
The latter two are related to two emission angles under which the light ray is sent to the receiver, i.e., the initial direction.

\subsection{Schwarzschild spacetime}
The metric in the usual coordinates is
\begin{multline}
	g = - \left(1- \frac{2m}{r}\right)c^2 dt^2 + \left(1- \frac{2m}{r}\right)^{-1} dr^2 \\ + r^2(d\vartheta^2 + \sin^2 \vartheta d\varphi^2) \, .
\end{multline}
We proceed to isotropic coordinates, which are better suited and allow for a straightforward post-Newtonian limit in the weak field approximation. 
In these coordinates, the metric reads
\begin{multline}
	g = -\left( \dfrac{1-\rho_s/\rho}{1+\rho_s/\rho}\right)^2 c^2 dt^2 \\ + \left( 1 + \rho_s/\rho \right)^4 \left( d\rho^2 + \rho^2 (d\vartheta^2 + \sin^2 \vartheta d\varphi^2) \right) \, .
\end{multline}
Here, $\rho \in (\rho_s,\infty]$ is the isotropic radial coordinate and $2 \rho_s = m = GM/c^2$ and $m$ is the only non-vanishing mass moment, the monopole, in natural units. An observer on the Earth's surface is located at some $\rho = \rho_0$ and the static redshift potential is 
\begin{align}
	e^{\phi} = \left( \dfrac{2-m/\rho}{2+m/\rho}\right) \, .
\end{align}
Level surfaces are topological spheres $\rho=\text{const.}$
The redshift of two static observers becomes
\begin{align}
	z + 1 = e^{\Delta \phi} = \dfrac{ (2-m/\rho_0)(2+m/\rho)}{(2+m/\rho_0)(2-m/\rho)} \, .
\end{align}
This expression allows to extract the weak-field limit as the contribution which is linear in $m$,
\begin{align}
	z = \Delta U / c^2 + \mathcal{O}(c^{-4}) \, , \quad U := - \dfrac{GM}{R} \, ,
\end{align}
where $R$ is the Euclidean radius.
Note that the first part of the above result generalizes also to more complicated Newtonian gravitational potentials in the post-Newtonian formalism such that, to first order, the redshift is proportional to $\Delta U$.

The Earth is rotating and, thus, there are Doppler contributions to the observed redshift.
An observer on the rotating Earth has a four-velocity
\begin{align}
	(\tilde{u}^\mu) = \tilde{u}^t(1,0,0,\tilde{v}^\phi) \, , \quad \tilde{v}^\phi=:\tilde{\Omega} = \text{const.}
\end{align}
and a (geodesic or non-geodesic) satellite is described by
\begin{align}
	(u^\mu) = u^t(1,v^i) \, .
\end{align}
Their redshift is
\begin{align}
	1 + z(t) 
	= e^{\Delta \phi} \dfrac{ \sqrt{1 - e^{-2\phi_0} \dfrac{\tilde{v}^2}{c^2} } }
{\sqrt{1 - e^{-2 \phi} \dfrac{v(t)^2}{c^2} }} \dfrac{1+ \frac{k_i(0)v^i(t)}{k_t}}{1+b\tilde{\Omega}} \, ,
\end{align}
with the constant impact parameter $b = k_\varphi / k_t$, and
\begin{align}
	\tilde{v}^2 &= \tilde{\Omega}^2 \rho_0^2 \left( 1 + \rho_s/\rho_0 \right)^4 \sin^2 \vartheta_0 \, , \quad \phi_0 = \phi(\rho_0) = \text{const.} \notag \\
	v^2 &= g_{ij}v^iv^j \, .
\end{align}
Introducing the new time scale $T = \exp(-\phi_0)t$, which is the proper time of static clocks at $\rho_0$ (on the non-rotating geoid), the redshift becomes
\begin{align}
	1 + z(T) 
	= e^{\Delta \phi} \dfrac{ \sqrt{1 - \dfrac{\tilde{V}^2}{c^2} } }{ \sqrt{1 - e^{-2 \Delta \phi} \dfrac{V(T)^2}{c^2} } } \dfrac{1+ \frac{k_i(0)V^i(T)}{k_T}}{ 1+b\tilde{\Omega} } \, ,
\end{align}
where $V^i = \exp(-\phi_0)v^i$ is the observed velocity on the new time scale.
If the satellite moves on a geodesic, we have
\begin{align}
	1+z(T) = e^{2\Delta \phi} \dfrac{E_T}{c^2} \dfrac{ \sqrt{1 - \dfrac{\tilde{V}^2}{c^2} } } {1+b\tilde{\Omega} } \left( 1 + \dfrac{k_i(0) V^i(T)}{k_T} \right) \, .
\end{align}
This relation can be used to create redshift models for geodetic satellites in the simple Schwarzschild spacetime.
Note that the EOP must be solved for the unknown components of $k_i$.

In general, the mass monopole can be determined from redshift measurements either between two Earth-bound observers or via Earth-satellite observations. 
In a very simple model, the satellite may move on a circular geodesic in the equatorial plane, leading to
\begin{align}
	z(b)= e^{2\Delta \phi} \dfrac{E_T}{c^2} \sqrt{ 1 - \dfrac{\tilde{V}^2}{c^2} }    \dfrac{\big( 1 + b \Omega \big)}{\big(1+b\tilde{\Omega} \big)} - 1 \, .
\end{align}
Note that the impact parameter $b$ is a function of the satellite's position. 
It can be related to the emission angle $\alpha$ w.r.t.\ the radial direction by
\begin{align}
	b = \dfrac{\tan \alpha \, \rho \left(1+\dfrac{m}{2\rho} \right)^3}{c \sqrt{1+\tan^2 \alpha} \left(1-\dfrac{m}{2\rho} \right)} \, .
\end{align}
Hence, the redshift is a function of $\alpha$. 
The angle $\alpha$ that belongs to the momentary satellite position is determined by the solution to the EOP.

In a second simple model, we might use radial signal transfer such that $(k_\mu) = (k_T,k_\rho,0,0)$ but do not require circular motion.
Then
\begin{align}
	1+z(T) = e^{2\Delta \phi} \dfrac{E_T}{c^2} \sqrt{1 - \dfrac{\tilde{V}^2}{c^2} } \left( 1 + \dfrac{k_\rho(0) V^\rho(T)}{k_T} \right) \, ,
\end{align}
where 
\begin{align}
	\dfrac{k_\rho(0)}{k_t} = \dfrac{\left(1+\dfrac{m}{2\rho} \right)^3}{\left(1-\dfrac{m}{2\rho} \right)} \, ,
\end{align}
and $V^\rho$ is to be determined from the geodesic equation.

For the mirror tracking in Schwarzschild spacetime, we assume the radial free-fall of a mirror to illustrate the idea. 
The result then is
\begin{align}
	1 + \dfrac{\nu_\text{em}}{\nu_\text{rec}} = \lim_{\Delta t_1 \to 0} \dfrac{\Delta T}{\Delta t_1} = 1 + \dfrac{v}{c} \left( 1 + \dfrac{2GM}{c^2 r} \right) + \mathcal{O}\left( c^{-4} \right) \, ,
\end{align}
where $v$ is the line-of-sight velocity.
Hence, observations of the two-way redshift allow us to deduce the velocity as done in conventional Doppler tracking, but also contain corrections due to the curvature (Shapiro delay).
Note that, in general, the velocity along the mirror's worldline will also change due to spacetime curvature.

\begin{figure}
	\centering
	\includegraphics[width=0.49\textwidth]{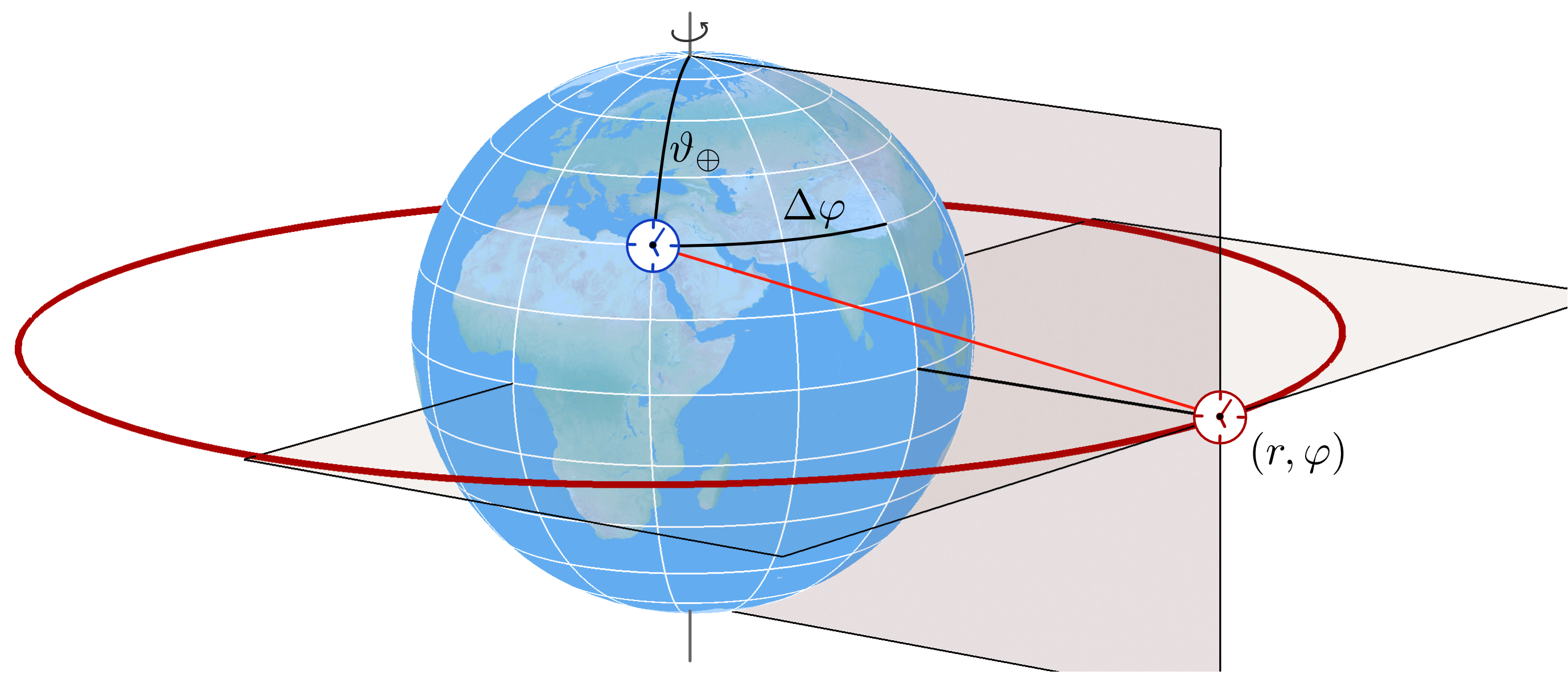}
	\caption{Sketch of signal exchange between an orbiting satellite and an Earth-bound clock.}
\end{figure}

\subsection{Kerr}
The Kerr metric in the well-known Boyer-Lindquist coordinates $(t,r,\vartheta,\varphi)$ reads
\begin{align}
	g 	&= - \left( 1-\dfrac{2mr}{\rho^2} \right) c^2 dt^2 + \dfrac{\rho^2}{\Delta} dr^2 + \rho^2 d\vartheta^2 \notag \\
		&+ \sin^2 \vartheta \left( r^2 + a^2 + \dfrac{2m r a^2\sin^2\vartheta}{\rho^2} \right) d\varphi^2 \notag \\
		&- \dfrac{4mra\sin^2\vartheta}{\rho^2} \, c \, dt d\varphi \, , 
\end{align}
where
\begin{align}
	\rho^2 = r^2 + a^2 \cos^2 \vartheta \, , \quad \Delta = r^2 + a^2 -2mr \, .
\end{align}
For the Kerr spacetime, there exists also a harmonic coordinate system, of which the non-rotation limit leads to the harmonic coordinates for the Schwarzschild spacetime.
Details on applications of the Kerr metric and properties such as its normal gravity field can be found in Ref.\ \cite{Kopeikin:2018} and references therein.
Here, we introduce new coordinates $\varphi' = \varphi - \omega t$, where $\omega$ is the angular velocity of an observer on the rotating Earth's surface that is thought of as being rigidly rotating.
Thus, 
\begin{align}
	g_{tt} \to g_{tt} + \omega^2 g_{\varphi\varphi} + 2 \omega g_{t\varphi} \, , \quad g_{t\varphi} \to g_{t\varphi} + 2 \omega g_{\varphi\varphi} \, .
\end{align}
In these coordinates, there is a redshift potential \cite{Philipp:2017,Philipp:2020}, as seen by the rigidly rotating observers on the Earth\footnote{These observers are at rest w.r.t.\ the new spatial coordinates.}, which is
\begin{align}
	e^{2\phi} & = 1-\dfrac{2mr}{\rho^2} + 
	4 \, \dfrac{\omega}{c} \,  \dfrac{amr\sin^2\vartheta}{\big( r^2+a^2\cos^2\vartheta \big)} \notag \\
	& \quad - \, \dfrac{\omega^2}{c^2} \,  
	\sin^2 \vartheta \left( r^2+a^2+\dfrac{2m r a^2\sin^2\vartheta}{r^2+a^2\cos^2\vartheta} \right) \, .
\end{align}
This redshift potential has two free parameters, $m$ and $a = J/(Mc)$, where $J$ is the angular momentum. 
Let an observer on the Earth's geoid be at position $\tilde{x} = \tilde{x}_0 = \text{const.}$ and make the proper time $T$ on the geoid a new coordinate time using ${T = \exp(-\phi(\tilde{x})) t =: \exp(-\phi_0) t}$.
The redshift w.r.t.\ any observer at position $x$ on the Earth's surface is given by
\begin{align}
	z + 1 = e^{\phi_0 - \phi(x)} \, .
\end{align}
Using the model of $\phi$, the two free parameters can be determined.
If the second clock moves on an orbiting satellite, Eq.\ \eqref{Eq:redshiftEarth} can be used to determine the potential.
If the satellite is on a geodesic, Eq.\ \eqref{Eq:redshiftEarthGeodesic} can be used instead.

Since the Kerr metric is a two-parameter family of solutions, it only allows one to i) either match the mass monopole and mass quadrupole to those of, e.g., the Earth at the cost of a mismatch of the spin dipole, or ii) match the mass monopole and spin dipole with the consequence that the mass quadrupole is far off; see also \cite{Kopeikin:2018}.
Thus, the Kerr spacetime is certainly not a good model for the Earth's exterior spacetime, instead it is preferable to consider a stationary metric of the form~\eqref{Eq:stat_metric} with a multipole expansion as described in the beginning of section \ref{sec:applications}.

\section{Summary and Outlook}
In this work, we have developed an exact general-relativistic framework for chronometric leveling on Earth and in space. Detailed analytical prescriptions of the methodology are provided for determining the stationary (gravitoelectric) gravitational potential while accounting for the motion of clocks. In particular, we derived analytical expressions for the redshift between arbitrary moving observers in stationary spacetimes and showed how clock comparisons can be used to determine the relativistic gravitational potential and, consequently, the multipole structure of the underlying spacetime.

A particularly elegant result is the neat factorization of the redshift into geometric, transversal-Doppler, and longitudinal-Doppler contributions. This provides a transparent description of how gravitational and kinematic effects enter clock comparisons and, importantly, establishes how the gravitational contribution can be extracted once the satellite state and signal propagation are properly modeled. An analogous factorization can also be achieved using energy conservation along a satellite's worldline, similar to the Newtonian energy formulation. We illustrated our general results for some specific configurations, involving two geodesic clocks, two stationary clocks, and a stationary Earth-bound clock compared with a moving satellite. In the weak-field limit, our results coincide with known notions and expressions, while the framework itself is formulated entirely within General Relativity without relativistic approximations beyond the assumption of stationarity.

For applications in, e.g., geodesy several aspects need further discussion. 
First, time-dependent gravitational effects, including tides and other variations of the Earth’s gravity field, have to be incorporated. As radiated energy due to gravitational wave emission can be safely neglected for the Earth, within the present framework such effects may be treated by allowing the multipole moments to vary adiabatically with time, as done in the conventional (Newtonian) framework. Second, the accuracy with which the satellite state, in particular its velocity, can be determined must be investigated carefully, since Doppler contributions can be substantially larger than the gravitational redshift. In this context two-way measurements should be investigated within the framework presented here. Likewise, an efficient solution and implementation of the emitter-observer problem either by analytical approximations or numerical simulations is mandatory; first steps towards a solution have been presented in \cite{Hackstein2025,Schlake:2025btk}.   

Satellite–to-satellite clock comparisons are particularly promising in the sense that the difficult to model atmospheric disturbances associated with ground-to-space links can be avoided with suitable constellations. They may therefore provide a stable complementary observable for gravity-field recovery once high performance space clocks become available. A constellation of satellites equipped with high-performance clocks could consequently enable a globally distributed chronometric data set. To quantify the achievable performance, numerical simulations for clock-equipped satellite constellations are currently under development \cite{Hackmann:2025ybh,Shabanloui2024}.

In this way, the framework for chronometry in space may develop into a complementary data channel for future gravity-field missions, providing a direct relativistic observable that can be combined with conventional geodetic measurements for improved modeling of the Earth’s gravitational field and geoid.

\begin{acknowledgments}

We gratefully acknowledge the Deutsche Forschungsgemeinschaft (DFG, German Research Foundation) under Germany’s Excellence Strategy – EXC-2123 QuantumFrontiers – 390837967, as well as within the CRC TerraQ – Project-ID 434617780 – SFB 1464 and the Research Unit RU 5456 Clock Metrology - Project-ID 490990195.
This work was also supported by the German Space Agency DLR with funds provided by the Federal Ministry of Economics and Technology (BMWi) under Grant No.\ DLR 50WM1547 and the DLR institute for Satellite Geodesy and Inertial Sensing.
\end{acknowledgments}



\bibliography{redshift}

\end{document}